# How and Why Inertial Mass and Gravitational Mass are Equal and Identical

*by*
*Roger Ellman*

Abstract

Present physics theory deems that the <u>inertial</u> mass characteristic of matter is the result of the interaction of that matter with a newly defined additional field called the Higgs Field after its principle researcher. Stated briefly, the Higgs field mechanism endows "gauge bosons" in a "gauge theory" with mass, through absorption of "Nambu–Goldstone bosons" arising in spontaneous "symmetry breaking".

Present physics theory deems that the <u>gravitational</u> mass characteristic of matter is the result of that matter "curving" or "warping" space and time according to the matter's presence through a mechanism not yet discovered nor defined.

Present physics recognizes that inertial mass and gravitational mass are equal as an empirical fact based on highly precise experiments. But, for example, the Earth in its orbit around the Sun experiences attraction toward the Sun involving its gravitational mass, $m_g$, and simultaneously experiences that attraction balanced by Earth's orbital centripetal force involving its inertial mass, $m_i$.

$$G \times \frac{M \times m_g}{R^2} = \frac{m_i \times v^2}{R}$$

That the matter of planet Earth is being endowed "gauge theory bosons with mass through absorption of Nambu–Goldstone bosons symmetry breaking in a Higgs Field" while simultaneously "warping or curving its region of space" is beyond the unreasonable and is simply inconceivable.

The resolution of this issue / phenomenon is presented in the following.

Roger Ellman, The-Origin Foundation, Inc.
320 Gemma Circle, Santa Rosa, CA 95404, USA
RogerEllman@The-Origin.org
http://www.The-Origin.org

## How and Why Inertial Mass and Gravitational Mass are Equal and Identical

by

*Roger Ellman*

Present physics theory deems that the <u>inertial</u> mass characteristic of matter is the result of the interaction of that matter with a newly defined additional field called the Higgs Field after its principle researcher. Stated briefly, the Higgs field mechanism endows "gauge bosons" in a "gauge theory" with mass, through absorption of "Nambu–Goldstone bosons" arising in spontaneous "symmetry breaking".

Present physics theory deems that the <u>gravitational</u> mass characteristic of matter is the result of that matter "curving" or "warping" space and time according to the matter's presence through a mechanism not yet discovered nor defined.

Present physics recognizes that inertial mass and gravitational mass are equal as an empirical fact based on highly precise experiments. But, for example, the Earth in its orbit around the Sun experiences attraction toward the Sun involving its gravitational mass, $m_g$, and simultaneously experiences that attraction balanced by Earth's orbital centripetal force involving its inertial mass, $m_i$.

$$G \times \frac{M \times m_g}{R^2} = \frac{m_i \times v^2}{R}$$

That the matter of planet Earth is being endowed "gauge theory bosons with mass through absorption of Nambu–Goldstone bosons symmetry breaking in a Higgs Field" while simultaneously "warping or curving its region of space" is beyond the unreasonable and is simply inconceivable. The mechanism for inertial mass and that for gravitational mass must be mutually consistent. Their equality is the result of their identity, the matter embodying them both, and doing so simultaneously, together, as one.

The following development is a simplified description of the mechanism of inertial mass in the force of electric field and the related mechanism of gravitational mass in the force of gravitation with their consequent equality and identity. The References provide the comprehensive, quantitative and analytical description and derivation of the effects.

### DEVELOPMENT OF THE UNIFIED FIELD

The following "thought experiments" develop the concept.

<u>Electric Field</u>

- Nothing can travel faster than the speed of light, $c$. Given two static electric charges separated and with the usual Coulomb force between them, if one of the charges is moved the change can produce no effect on the other charge until a time equal to the distance between them divided by $c$ has elapsed.

- For that time delay to happen there must be something flowing from the one charge to the other at speed $c$ and the charge must be the source of that flow.

  The Coulomb effect is radially outward from the charge, therefore every charge must be propagating such a flow radially outward in all directions from itself, which flow must be the "electric field".

## Unification of Fields

- Except for the kind of field, all of the preceding applies in the same way and with the same conclusions for gravitational field as for electric field.

- Therefore, either a particle that exhibits both such fields, as for example a proton or an electron, is a source of two separate and distinct flows, one for each field, or there is only a single flow which produces both effects, electric and gravitational.

- The only reasonable conclusion is that electric and gravitational field are different effects of the same sole flow from the source particles.

## Sources & Their Decay

- The flow is not inconsequential. Rather, it accounts for the forces, actions and energies of our universe.

- For a particle to emit such a flow the particle must be a source of whatever it is that is emitted outward. The particle must have a supply of it.

- The process of emitting the flow from a particle must deplete the supply resource for the particle's emitting further flow, must use up part of its supply, else we would have something-from-nothing and a violation of conservation.

    It must be concluded that an original supply of that which is flowing came into existence at the beginning of the universe and has since been gradually being depleted at each particle by its on-going outward flow.

## The Beginning

- Before the universe began there was no universe. Immediately afterward there was the initial supply of medium to be propagated by particles. How can one get from the former to the latter while: (1) not involving an infinite rate of change, and (2) maintaining conservation ?

    The only form that can accommodate the change from nothing to something in a smooth transition without an infinite rate of change is the oscillatory form of equation *(1a)*, below.

*(1a)* $\quad U_0 \cdot [1 - \cos(2\pi \cdot f \cdot t)]$

    The only way that such an oscillation can have come into existence without violating conservation is for there simultaneously to have come into existence a second oscillation, the negative of equation *(1a)* as in equation *(1b)*.

*(1b)* $\quad -U_0 \cdot [1 - \cos(2\pi \cdot f \cdot t)]$

    That is, the two simultaneous oscillations must have been such as to yield a net of nothing, the prior starting point, when taken together.

## That Which is Flowing

- The flow is a property of contemporary particles. Those particles are evolved successors to the original oscillations with which the universe began. Then, that which is flowing is the same original primal "medium", the substance of the original oscillations at the beginning of the universe.

    Since each flow is flowing outward from the myriad particles of the universe simultaneously and that flow is passing through myriad others of those particles' flows without untoward interference, the flow "medium" must be extremely intangible for all of that to take place, any one particle's flow flowing largely freely through that of other particles. It must be as intangible as -- well -- "field".

The Oscillatory Medium Flow

- The initial medium supply of each particle, each being a direct "descendant" of the original oscillation at the universe's beginning, must be oscillatory in form per equations *(1)*. Therefore the radially outward flow from each particle is likewise an oscillatory medium flow of the form of equations *(1)*.

  The flow is radially outward from the particle, therefore, the oscillation of the medium supply of each particle is a spherical oscillation. The particle can also be termed a *center-of-oscillation*, or *center*, which terms will also be used.

- The amplitude, $U_0$, of the *[1 – Cosine]* form oscillation is the amplitude of the flow emitted from the source particle, which flow corresponds to the electric field. Thus the oscillation amplitude must be the charge magnitude of the source particle -- the fundamental electric charge, $q$, in the case of the fundamental particles, the electron and the proton.

  Then, the conservation-maintaining distinction of amplitude $+U_0$ versus amplitude $-U_0$ must be the positive / negative charge distinction.

- The frequency, $f$, of the *[1 – Cosine]* form oscillation must then correspond to the energy and mass of the source particle, that is the energy of the oscillation is $E = h \cdot f$ and the mass is $m = E/c^2 = h \cdot f / c^2$.

## *MASS AND COULOMB REPULSION OF LIKE CHARGES*

Just as the *[1 – Cosine]* type oscillatory wave of propagated medium is the field, so the source of that propagation, the source itself from which the propagation is emitted radially outward in all directions must embody the charge and the mass, that is, the matter of the "particle" involved.

The effect "force" is, then, the result of the waves propagated by a *center-* or *centers-of-oscillation* arriving at and interacting with an encountered *center*, the one upon which the force is exerted. (In the following discussion, *centers-of-oscillation* will be referred to as the "*source*" *center* and the "*encountered*" *center*). Of course every *center* is continuously in both roles.

The effect of an individual cycle of wave encountering a center is the delivery to the center of an *impulse,* a *force × time,* an amount of *momentum change*, in the direction away from the source center, the direction of repulsion. The amount of impulse in the wave is, of course, proportional to the amplitude of the wave. It is that amount, that amplitude, which decreases as the square of the distance from the source center because it becomes spread over a greater area. The overall stream of successive wave cycles carries the impulse of one wave times the repetition rate, the frequency, of the waves.

Newton's Law,

*(2)*    Force = Mass × Acceleration

can be restated as

*(3)*    Acceleration Resulting = Force Applied × $\dfrac{1}{\text{Mass}}$

and in that form is a more natural statement since force is the cause and acceleration the effect. This translates in terms of waves and centers into

*(4)*    $\begin{bmatrix} \text{Acceleration} \\ \text{Resulting} \end{bmatrix} = \begin{bmatrix} \text{Wave} \\ \text{Impulse} \end{bmatrix} \times \begin{bmatrix} \text{Responsiveness} \\ \text{of the Center} \end{bmatrix}$

or, more succinctly,

   Acceleration = Wave × Responsiveness.

The responsiveness depends upon the encountered center cross-section, the effective "target" area that the encountered center has for intercepting incoming waves. Of the total wave traveling outward from the source center, the only part that interacts with another center is the part that encounters the center, that is intercepted by the encountered center.

The source center's radially outward flowing wave front is a sphere of steadily increasing radius, of steadily increasing total surface area. If the encountered center is sufficiently distant from the source center then the portion of the source center's flow that encounters the encountered center is a small part of the source center's total spherical propagation. It is effectively a plane wave; that is, the part of the wave intercepted by the encountered center is essentially a flat wave front of which every part travels parallel to the center part.

A center of smaller cross-section (smaller "target" = lesser responsiveness) is of greater inertial mass, all other effects being equal, and requires greater arriving wave amplitude to experience a specified change in motion than does a center of larger cross-section. Cross-section is a matter of size, that is it is proportional to the area of interception of the incoming wave front. The encountered center being a spherical oscillation the cross-section is the area of a circle perpendicular to the direction of travel of the wave front as it encounters the center.

In order to account for the action of waves on an encountered center relating only to that portion of the total wave front intercepted by the encountered center, the incoming wave must be expressed in terms of "Incoming Wave Impulse per Unit Area". That is, the intercepted wave impulse, the "Wave" of equation *(4)* is

*(5)*
$$\text{Wave} = \frac{\text{Total Propagated Wave Impulse of Source Center}}{\substack{\text{Total Spherical Area of Source Wave at} \\ \text{Distance Encountered Center is from Source}}}$$

$$= \text{Wave Potential Impulse per Unit Area there}$$

so that upon being multiplied by the cross-sectional area at the encountered center the units of area are cancelled and the resulting quantity is wave impulse (as measured at and as intercepted by the encountered center). The division by the area of a sphere is the essence of the inverse square law, of course.

### *OPPOSITE CHARGES COULOMB ATTRACTION*

For each center emitting its outward medium flow, emitting outward impulse, conservation requires an equal opposite reaction, an inward impulse toward the center of the emitting center-of-oscillation. Since the outward medium flow is radially outward in all directions equally the net effect of the inward reactions is null.

If, however, a center is encountered by an incoming opposite amplitude (negative to positive or positive to negative) wave, that wave offsets part of the outward flow's inward reaction on the encountered center in the direction of the incoming flow. But the corresponding inward reaction remains at its full amplitude on the side of the encountered center directly opposite where the incoming opposite amplitude wave is impacting.

The result is a net impulse imbalance on the encountered center, an imbalance directed toward the source center, and of magnitude determined by the incoming wave. That is, the Coulomb Effect between opposite charges is one of attraction as compared to the repulsive effect between like charges.

### *GRAVITATIONAL ATTRACTION*

The same arriving wave has a second, much smaller effect on the encountered center. The encounter of two medium flows directly "head on" results in slowing of the otherwise normal at *c* velocity of propagation of each in proportion to the density or concentration of each flow.

ε, μ, and the Speed of Propagation

A brief consideration of an electrical analog to medium wave propagation is necessary. A transmission line is an electrical device, such as coaxial cables and wire pairs found in radio, and video systems, for transmitting oscillatory electrical energy from one place to another. When electrical signals are introduced at one end of such a line there is a finite speed of travel of the electrical effects along the line.

The reason is that the line has electrical inductance and capacitance distributed over its length and each introduces a delay in fully responding to an input signal.

The speed of propagation along the transmission line, the speed of propagation through a medium of distributed inductance and capacitance of values per unit length $L_p$ and $C_p$, is

(6) $$v = \frac{1}{\sqrt{L_p \cdot C_p}}$$

This same result applies to light, a propagation through space, which has inductance per unit length of $\mu_0$ and capacitance per unit length of $\varepsilon_0$, giving the speed of light as

(7) $$c = \frac{1}{\sqrt{\mu_0 \cdot \varepsilon_0}}$$

The original medium flow at the beginning or the universe was into empty space, complete nothing, and the medium flow of today's particles is the successor to that original flow. The medium flowing through empty space between particles, flows likewise at the same $c$ as above due to an effective $\mu_0$ and $\varepsilon_0$ combination as above. But, where do its $\mu_0$ and $\varepsilon_0$ come from; how does empty "free space", the nothing that was before the universe began, have those characteristics ?

It cannot and does not. Until medium flow appears the "free space" is absolute nothing, the non-existence of before the origin of the universe. Clearly, it must be the medium itself, the only non-nothing material reality, that is the cause of $\mu_0$ and $\varepsilon_0$.

The amount of medium at a particular location determines, the value of $\mu_0$ and $\varepsilon_0$ at that location. That quantity, the medium amount is a scalar quantity, one having magnitude but not an associated direction. The medium flow is a vector quantity, having magnitude and direction. Medium from a single source, diffusing into greater spherical volumes in space maintains constant speed of propagation, $c$, because the ratio of the medium amplitude to the $\mu_0$ and $\varepsilon_0$ remains constant

But, when medium propagating from one source passes through the same space as medium from some other source, depending on the orientation of the flows the effect can be a reducing of the speed of propagation, the $c$ of both medium flows.

If two mutually encountering medium flows are traveling in the same direction their parameters all combine. The combined flow balances to the combined $\mu_0$ and $\varepsilon_0$. Just as medium from a single source, diffusing into greater spherical volumes in space maintains constant speed of propagation, $c$, because the ratio of the medium amplitude to the $\mu_0$ and $\varepsilon_0$ remains constant, so two medium flows in the exact same direction have, combined, the same ratio of amplitude to $\mu_0$ and $\varepsilon_0$ as do their individual flows taken separately.

But if those two flows are in exactly opposite directions, through each other, a different result occurs. The scalar $\mu_0$ and $\varepsilon_0$ of the individual flows still combine to produce new somewhat greater values. But, the flows are vectors inverse-square diffusing in opposite directions and there is no valid net resultant. Each flow acts and must act as if alone, propagating independently in its own direction.

Thus, for medium flows propagating in opposite directions through each other there is for each flow an increase in the $\mu_0$ and $\varepsilon_0$ that it must address for which it has available only the same, now proportionally less sufficient, amplitude of its own to drive itself. Each flow encounters a partially reduced value of $c$. Each is effectively slowed by the other.

If the two medium flows encounter each other neither exactly in the same nor opposite directions, they can be resolved into mutually parallel and perpendicular components. The parallel components fit one of the above two cases with the magnitudes being those of the components

Gravitational Attraction

Gravitational attraction is caused by the above slowing of the speed of flow of medium waves. The incoming waves from the source center flow directly opposed to the outward waves being emitted by the encountered center and consequently slow that emission. The emitted propagation being slowed carries reduced impulse.

Since the encountered center's outward propagation impulse is reduced in the direction of the source center, its equal opposite reaction inward, away from the source center is reduced. But the corresponding inward reaction remains at its full amplitude on the side of the encountered center directly opposite where the incoming wave produced the slowing.

The result is a net impulse imbalance on the encountered center, an imbalance directed toward the source center, and of magnitude determined by the incoming wave's slowing effect.

## CONCLUSION

Thus gravitational attraction is caused by the same mechanism in the encountered center as Coulomb attraction between opposite charges.

Both masses depend on the "target" that the encountered center is for the incoming waves from the source center.

All centers are always simultaneously in both roles: source and encountered.

The magnitudes of the Coulomb and Gravitational effects are different because they employ different mechanisms of affecting the encountered center with the same medium wave flow.

Inertial mass and gravitational mass are not merely equal; <u>they are identical</u>.

See the following references for the comprehensive, quantitative, analytical development of all the above.